\documentclass[a4paper]{iopart}
\pdfoutput=1
\usepackage{amsmath}
\usepackage{amssymb}
\usepackage{graphicx} 
\usepackage{hyperref}
\usepackage{color}

\newcommand{\ket}[1]{\left|#1\right\rangle}
\newcommand{\bra}[1]{\left\langle #1\right|}
\newcommand{\ud}{\mathrm{d}}

\newcommand{\mean}[1]{\left\langle #1\right\rangle}

\newcommand{\nep}{\textrm{e}}
\newcommand{\atan}{\operatorname{atan}}

\begin{document}

\title{Floquet resonances close to the adiabatic limit and the effect of dissipation}

\author{Angelo Russomanno$^{1,2}$ and Giuseppe E. Santoro$^{2,3,4}$}

\address{$1$ Scuola Normale Superiore, Piazza dei Cavalieri 7, I-56127 Pisa, Italy}
\address{$2$ International Centre for Theoretical Physics (ICTP), Strada Costiera 11, I-34151 Trieste, Italy}
\address{$3$ CNR-IOM Democritos National Simulation Center, Via Bonomea 265, 34136 Trieste, Italy}
\address{$4$ SISSA, Via Bonomea 265, 34136 Trieste, Italy}


\begin{abstract}
{We study the approach to the adiabatic limit in periodically driven systems. 
Specifically focusing on a spin-1/2 in magnetic field we find that, when the parameters of the Hamiltonian lead to a quasi-degeneracy in the Floquet spectrum, 
the evolution is not adiabatic even if the frequency of the field is much smaller than the spectral gap of the Hamiltonian. 
We argue that this is a general phenomenon of periodically driven systems. 
Although an explanation based on a perturbation theory in $\omega_0$ 
cannot be given, because of the singularity of the zero frequency limit,  
we are able to describe this phenomenon by means of a mapping to an extended Hilbert space, in terms of resonances of an effective two-band Wannier-Stark ladder.
Remarkably, the phenomenon survives in presence of dissipation towards an environment 
and can be therefore easily experimentally observed. }
\end{abstract}

\section{Introduction}
The adiabatic theorem~\cite{Messiah:book} is one of the cornerstones of the quantum mechanics: it is deeply linked to the concept of geometric phases~\cite{Berry}, 
to the Kibble-Zurek scaling phenomenon~\cite{Polkovnikov_RMP11,Zerek_PRL05,Dziarmaga_AP10}, and to Quantum Annealing \cite{Finnila_CPL94,Kadowaki_PRE98,Brooke_SCI99,Santoro_SCI02} {\em alias} Adiabatic Quantum Computation \cite{Farhi_SCI01}, 
to name just a few applications. 
Recently, there has been interest in extensions of the adiabatic theorem to many-body periodically driven systems whose parameters are slowly varied in time, 
especially in connection with non-equilibrium quantum phase transition~\cite{Breuer_ZPD89,Eckardt_PRL051,Polko_adiab,Lorenzo,Emanuele_arXiv15,Emanuele_arXiv16}. 
In this case, the properties of the Floquet states --- the eigenstates of the one-period evolution operator ---, and the associated quasi-energies, are crucial. 

Ref.~\cite{Marzlin_PRL04} pointed out that the traditional condition for the validity of the adiabatic approximation ---
the Hamiltonian being slowly changing with respect to the inverse squared gap --- is not sufficient to provide adiabaticity, 
although necessary~\cite{Tong_PRL10}, and a system can deviate from its adiabatic regime even if it is driven very slowly. 
Due to the very widespread applications of the adiabatic theorem, it is very important to understand when these deviations occur and to which extent 
they affect the dynamics of the system. 
For instance, it is interesting to understand if they can be still observed in an experimental context where there are unwanted interactions with the environment and decoherence.
Here we choose to focus on periodically driven systems, due to their importance in the recent theoretical and experimental developments of quantum mechanics (see Refs.~\cite{Grifoni_PR98,Eckardt_RMP_17} for a review).  
We select the simplest and most paradigmatic one: a single spin $1/2$ in a time-periodic magnetic field, which will provide us an understanding relevant also for more complex cases. 
In agreement with Refs.~\cite{Marzlin_PRL04,Tong_PRL10} we observe that there are some frequencies where the system should be adiabatic because the driving frequency $\omega_0$ is
much smaller than the minimum gap of the instantaneous Hamiltonian but there is no adiabaticity at all: observables strongly deviate from their adiabatic value and perform large quantum beats reminiscent of Rabi oscillations. 

Our main improvement is that we apply Floquet theory to this problem and we are able to understand the condition which $\omega_0$ has to obey in order to provide these anomalous 
non-adiabaticities. We find that the anomalous non-adiabatic frequencies correspond to quasi-degeneracies of the Floquet spectrum. 
At these quasi-degeneracies, an integer number of frequencies of the external driving approximately matches the difference between two Floquet quasi-energies, like a multi-photon resonance. 
Therefore the condition ``$\omega_0$ much smaller than the minimum gap'' has to be supplemented by ``$\omega_0$ away from any Floquet quasi-degeneracy'' in order to have adiabaticity. We believe that this is true in general for periodically driven systems. 

Looking more carefully at the properties of Floquet quasi-degeneracies in our problem, we meet an infinite number of them if we tend towards the limit of zero driving-frequency: the vanishing frequency limit cannot be interpreted as a perturbation expansion in $\omega_0$. 
Remarkably, we can give an interpretation of this phenomenon by means of the extended Hilbert space representation~\cite{Shirley_PR65,Sambe_PRA73,Hausinger_PRA10}. 
Using an analysis similar to that of the multi-frequency case considered in Ref.~\cite{Halperin_arxiv}, we see that the dynamics can be mapped to that of a particle in a two-band 
Wannier-Stark ladder~\cite{mendez_PT81,Wannier:book} which we will refer to as ``Floquet-Stark ladder''. 
We will show that the case of a frequency $\omega_0$ far from any quasi-degeneracy corresponds to the absence of resonances between the levels of the Floquet-Stark ladder: 
this gives rise to Bloch oscillations~\cite{Grosso:book} in the ladder, which correspond to the adiabatic evolution of the energy. 
On the contrary, the quasi-degenerate case corresponds to a state in the lower-band-manifold of the Floquet-Stark ladder which becomes resonant with another level in the upper-band-manifold: 
the resulting Rabi oscillations give rise to the large non-adiabatic oscillations of the energy.

In order to understand to which extent the Floquet resonances affect the behaviour of the system, we consider also the effect of dissipation. 
This is also important for experimental realizations where the coupling with the environment cannot be avoided. 
We discuss this problem both from a classical and a quantum perspective. Indeed, the non-adiabatic beatings are observed also in the purely classical precession equation 
obeyed by the expectations of the spin components, so one can imagine to perform an experiment with a bulk classical magnetization. 
The dissipation in this case is described by the Landau-Lifshits-Gilbert equation~\cite{LLG:review} and we see that, before a stationary condition is attained, 
the non-adiabatic beatings can be observed in the dynamics of the system. 
From a quantum point of view, we consider a single spin coupled to a thermal bath: this setting is interesting for superconducting qubit experiments~\cite{Clarke_nat08,Devoret_sci13}.
Expanding the Bloch-Refdield master equation in the Floquet basis we study the properties of the periodic stationary condition attained by the system. 
We see that the energy in the steady state shows marked peaks at the Floquet quasi-degeneracies: the effect of these non-adiabaticities extends far beyond the unitary-evolution 
setting where we have described them.
%
\section{Floquet quasi-degeneracies and adiabaticity breaking in the unitary dynamics}

\begin{figure}
\begin{center}
   \includegraphics[width=8.3cm]{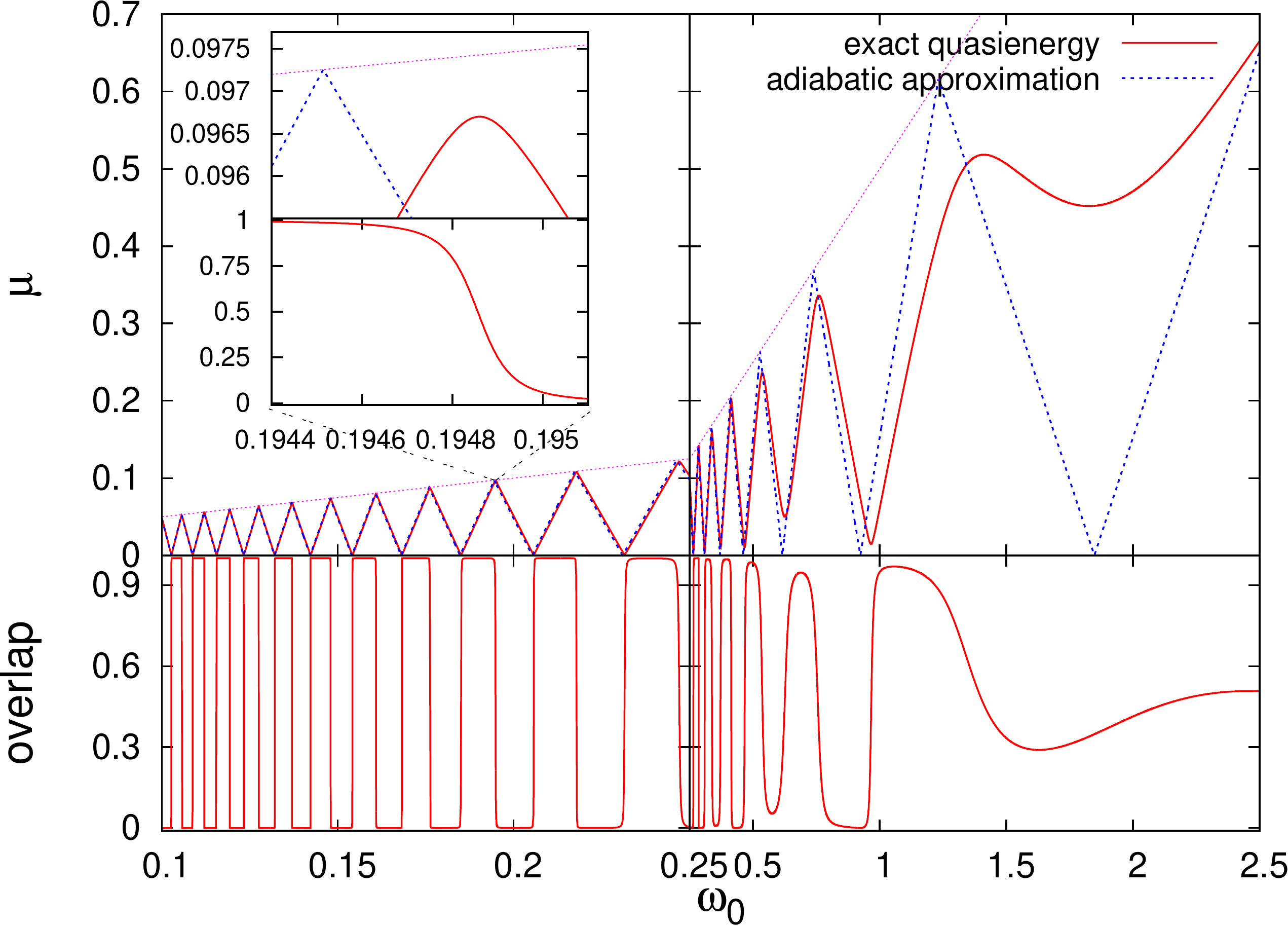}
\end{center}
\caption{(Upper panel) Approximate adiabatic (blue dashed line) and exact (red line) positive Floquet quasi-energies vs. $\omega_0$ in the first Brillouin zone.
 When two different quasi-energies are quasi-degenerate, the second quantity almost equals 0 or $\omega_0/2$, as explained in the main text.
(Lower panel) The square overlap $\left|\left\langle\Phi^{+}(0)\right.\ket{{\rm g}(0)}\right|^2$ of one of the Floquet modes at time 0 with the corresponding ground state of the Hamiltonian vs. $\omega_0$: notice 
 the switchings between 0 and 1 at the quasi-degeneracies which can be better seen in the inset. We take $\Delta=\epsilon=1$ and $A=2$.}
\label{floq_qeo:fig}
\end{figure}

A central tool for our discussion is the Floquet theory \cite{Shirley_PR65,Sambe_PRA73,Grifoni_PR98}. 
This theory states that, given a time-periodic Hamiltonian $\hat{H}\left(t\right)=\hat{H}\left(t+\tau\right)$, there exists a basis of solutions of the Schr\"odinger equation (Floquet states) 
which are periodic up to a phase
\begin{equation}
  \ket{\Psi_\alpha\left(t\right)}=\nep^{-i\mu_\alpha t}\ket{\Phi_\alpha\left(t\right)}\,.
\end{equation}
The periodic states $\ket{\Phi_\alpha\left(t\right)}=\ket{\Phi_\alpha\left(t+\tau\right)}$ are called Floquet modes, the real quantities $\mu_\alpha$ are called Floquet quasi-energies. 
The analogy with the Bloch theory for particles in crystalline potentials is clear:
the Floquet states are the analogue of Bloch waves and the quasi-energies are the analogue of quasi-momenta. 
We focus our discussion on the case of a spin in a time-periodic magnetic field
\begin{equation} \label{Hamiltonian:eqn}
  \hat{H}\left(t\right)= -{\bf B}(t) \cdot \hat{\boldsymbol{\sigma}}
\end{equation}
where ${\bf B}(t)=-(0,\Delta,\epsilon+A\cos(\omega_0 t))$ and $\hat{\sigma}^j$ are the Pauli matrices. 
To obtain the Floquet modes and quasi-energies, it is enough to know the time evolution operator for $0<t<\tau$~\cite{Grifoni_PR98} (here $\tau=2\pi/\omega_0$ is the period). 
Like the quasi-momenta of Bloch waves, the quasi-energies are defined up to translations of $\omega_0$: this implies that two quasi-energies are degenerate if they differ 
by an integer number of $\omega_0$. 
Such a translation symmetry allows therefore to define a Brillouin zone (BZ) structure in quasi-energy: we focus on the first BZ which lies 
between $-\omega_0/2$ and $\omega_0/2$. 
In the problem we are focusing on, Eq.~\eqref{Hamiltonian:eqn}, there are two distinct quasi-energies; they are equal in modulus but opposite in sign because the Hamiltonian has 
vanishing trace~\cite{Russomanno_PRL12}. 
In Fig.~\ref{floq_qeo:fig} (upper panel) we plot the positive quasi-energy vs $\omega_0$. 
We plot for comparison also the adiabatic approximation~\cite{Russomanno_PRB11} for the quasi-energy $\mu_{\rm ad}^e=\frac{1}{\tau}\int_0^\tau E_{\rm e}(t)\ud t$ 
(where $E_{\rm e}(t)$ is the energy of the instantaneous excited state at time $t$): we see that the adiabatic approximation becomes better for smaller values of $\omega_0$, as expected. 
We notice that in many points the quasi-energies approach 0 or $\pm\omega_0/2$. 
These are quasi-degeneracy points where two quasi-energies, up to translations of $\omega_0$, are almost equal. 
For small values of the frequency, the corresponding adiabatic approximations for the quasi-energies have exact degeneracies for values of $\omega_0$ near the quasi-degeneracies 
(see the upper inset in Fig.~\ref{floq_qeo:fig}). 
Considering lower and lower frequencies, the quasi-degeneracies become infinitely dense but they do not disappear. 
On the opposite, they become vanishingly narrow and -- anticipating a little bit -- the Floquet modes exchange each other on a vanishingly thin frequency range 
(see the lower panel of Fig.~\ref{floq_qeo:fig}). 
Zero frequency is an accumulation point of such switchings: the limit $\omega_0\to 0$ is singular~\cite{Avron_JPA99} and non-perturbative. 

\begin{figure}[ht!]
\begin{center}
   \includegraphics[width=8.3cm]{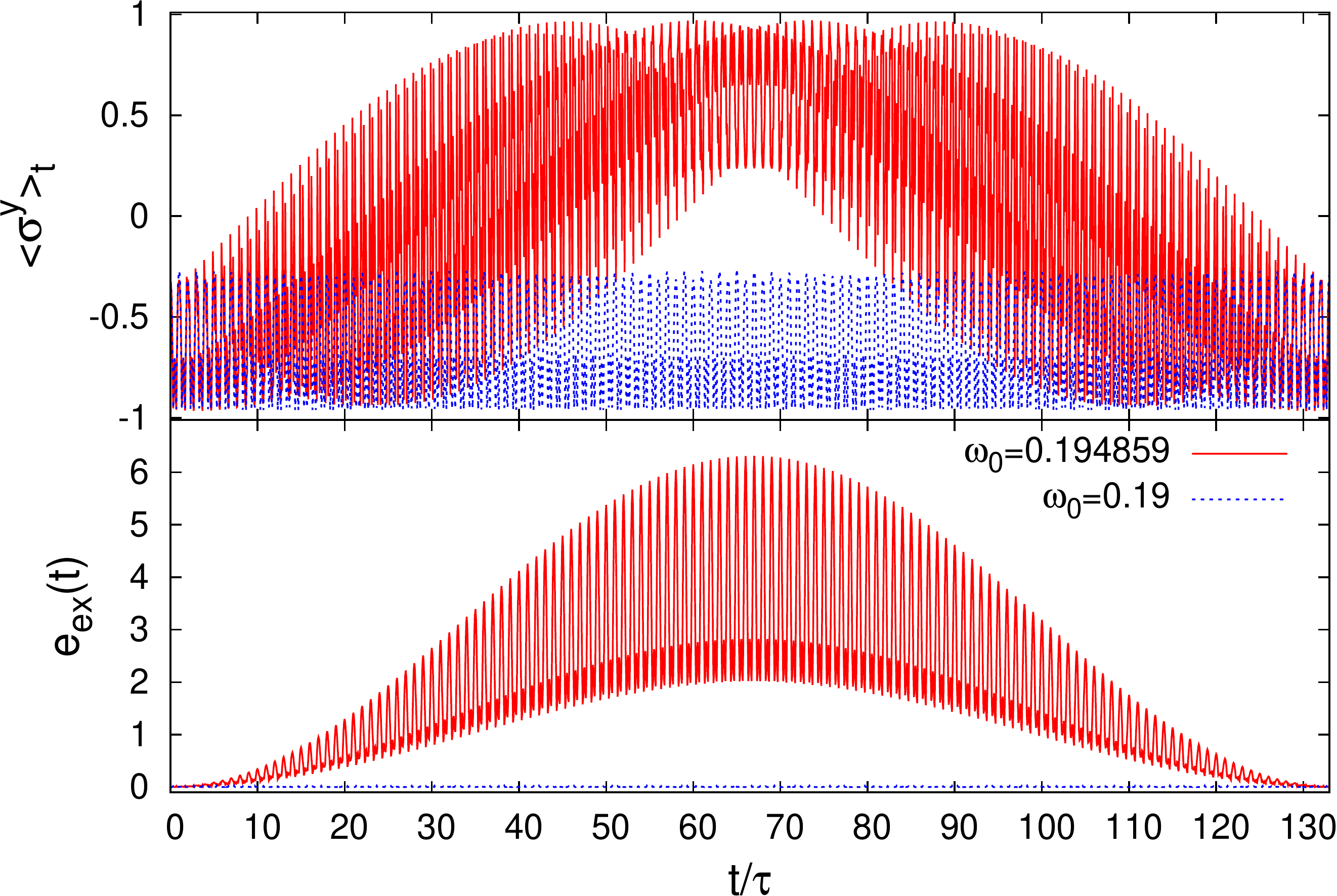}
\end{center}
\caption{
The expectation $\mean{\hat{\sigma}^y}_t$ (upper panel) and the excitation energy (lower panel) vs. the number of periods $t/\tau$ for two different frequencies 
($\omega_0=0.19$ and $\omega_0=0.194859$) both in the adiabatic limit $\omega_0\ll \min_\tau E_{\rm gap}\simeq2.8$. 
For $\omega_0=0.19$ (blue curves) the results are indistinguishable from the adiabatic prediction. 
For $\omega_0=0.194859$ we are at a quasi-degeneracy of the Floquet spectrum (see upper inset of Fig. \ref{floq_qeo:fig}) 
and we see a strong violation of adiabaticity and large-amplitude long-period beatings of the observables (red curves). (Numerical parameters: $\Delta=\epsilon=1$ and $A=2$.)
}
\label{time_unitary:fig}
\end{figure}

When the frequency $\omega_0$ is at a quasi-degeneracy there is a violation of adiabaticity, however small is the frequency. 
In order to understand this point, we focus on a frequency 
$\omega_0$ small enough to make the adiabaticity condition (see for instance~\cite{Rigolin_PRA08,Messiah:book}) valid:
\begin{equation} \label{adiabone:eqn}
  \omega_0\ll\min_{t\in[0,\tau]}E_{\rm gap}\left(t\right) \;.
\end{equation}
(Here $E_{\rm gap}\left(t\right)=E_{\rm e}\left(t\right)-E_{\rm g}\left(t\right)$ is the difference of the excited and ground energy eigenvalues at time $t$.) 
As discussed before, the implication of adiabaticity from this condition has been challenged in Ref.~\cite{Marzlin_PRL04} and then Ref.~\cite{Tong_PRL10} has shown that it is only necessary: we are going to study when this condition fails to provide adiabaticity.
If the spin dynamics is adiabatic, the expectation at time $t$ of the operator $\hat{\boldsymbol{\sigma}}$,
call it $\mean{\hat{\boldsymbol{\sigma}}}_t$, is always parallel to the instantaneous value of the magnetic field ${\bf B}(t)$, and time-periodic, with period $\tau$. 
In the upper panel of Fig.~\ref{time_unitary:fig}, the blue curve shows $\mean{\hat{\sigma}^y}_t$ for a small $\omega_0$ far from any quasi-degeneracy: we can see that 
the predictions based on the adiabatic theorem are perfectly verified in this case. 
On the opposite, changing a little bit $\omega_0$ so that Eq.~\eqref{adiabone:eqn} is still valid but there is a quasi-degeneracy, adiabaticity is destroyed, 
as we can see in the red curve of the upper panel of Fig.~\ref{time_unitary:fig}. 
This time trace shows a strong deviation from adiabaticity of $\mean{\hat{\sigma}^y}_t$ in the form of large beatings which last many driving periods $\tau$. 
The contrast is even more striking in the lower panel of Fig.~\ref{time_unitary:fig} where we plot the excitation energy defined as
\begin{equation}
  e_{\rm ex}(t)=\bra{\Psi\left(t\right)}\hat{H}\left(t\right)\ket{\Psi\left(t\right)}-E_{\rm g}\left(t\right)\,.
\end{equation}
%
%
For a perfectly adiabatic evolution, this quantity is always vanishing: this fact is verified for a frequency low and far from any Floquet quasi-degeneracy ($\omega_0=0.19$ in the plot). 
On the contrary, at a Floquet quasi-degeneracy the dynamics is non-adiabatic: after many periods $e_{\rm ex}(t)$ starts deviating from 0 and makes large oscillations between 0 
and its maximum value in the form of large beatings. 
It is important to remark that these beatings have a frequency corresponding to the gap in the Floquet spectrum at the quasi-degeneracy (we exploit the invariance of the quasi-energies under translations of $\omega_0$ in order to consider the minimum gap over all the possible relative $\omega_0$-translations of the levels): we will give a better interpretation of this point later. 
At a Floquet quasi-degeneracy, we have that an integer number of frequencies of the driving approximately matches the gap between two quasi-energies (chosen in different Brillouin zones); 
therefore we have a multi-photon resonance which physically justifies the observed beatings. 
%
We notice that the average values of the spin components obey a classical precession equation
\begin{equation}  \label{class}
  \frac{d}{dt}\mean{\hat{\boldsymbol{\sigma}}}_t=-2\mean{\hat{\boldsymbol{\sigma}}}_t\times{\bf B}(t),\,
\end{equation}
so these beatings can be observed not only for quantum spins but also for classical magnetic moments, for example in NMR experiments. 

\section{Adiabaticity breaking and properties of the Floquet modes}

We find that it is possible to describe the violation of adiabaticity at the quasi-degeneracy in terms of Floquet modes and quasi-energies in the low frequency limit. 
As already mentioned, in the limit of vanishing $\omega_0$, the quasi-energies are the average over one period of the instantaneous eigenenergies~\cite{Russomanno_PRB11}
\begin{equation} \label{adiabbondio:eqn}
  \mu_{\rm ad}^{e/g} = \frac{1}{\tau}\int_0^\tau E_{e/g}(t)\ud t\,,
\end{equation}
up to translations of $\omega_0$ (these are the adiabatic quasi-energies -- see Fig. \ref{floq_qeo:fig}) and the Floquet modes become the instantaneous eigenstates of the Hamiltonian $\left\{\ket{{\rm e}(t)},\,\ket{{\rm g}(t)}\right\}$. 
The reason for this fact is as follows: in the adiabatic limit the instantaneous eigenstates of the Hamiltonian $\left\{\ket{{\rm e}(t)},\,\ket{{\rm g}(t)}\right\}$ solve the Schr\"odinger equation 
and are periodic up to a phase $\mu_{\rm ad}^{\pm} t$, where $\mu_{\rm ad}^{\pm}$ is given by Eq.~\eqref{adiabbondio:eqn} in the limit 
$\tau\to\infty$.~\footnote{For this specific form of driving the Berry phase~\cite{Berry} $\gamma_{\pm}$ is vanishing. 
In general, the Berry's phase gives a contribution $\gamma_{\pm}/\tau$ to the adiabatic quasi-energies, which vanishes for $\tau\to\infty$.} 
Therefore, in the adiabatic limit, the square overlap $\left|\left\langle\Phi^{\pm}(0)\right.\ket{{\rm g}(0)}\right|^2$ of each Floquet mode at time 0 with the $t=0$-ground-state is 1 or 0.
In the lower panel of Fig.~\ref{floq_qeo:fig} we plot this overlap for $\ket{\Phi^{+}(0)}$ versus $\omega_0$: at low but non-vanishing frequencies we see that it is 1 or 0 when $\omega_0$ 
is far from any quasi-degeneracy. Around a quasi-degeneracy, the adiabatic prescription is not obeyed and the Floquet modes are different from the eigenstates of the Hamiltonian: crossing 
the quasi-degeneracy, the overlap $\left|\left\langle\Phi^{+}(0)\right.\ket{{\rm g}(0)}\right|^2$ goes continuously from 0 to 1 
(the Floquet states exchange each other -- see the inset of Fig. \ref{floq_qeo:fig}).  
At a quasi-degeneracy, therefore, the initial ground state is a superposition of two Floquet states which have a $2\pi/\omega_0$-periodic part and a modulation with slightly different 
quasi-energies modulo translations of $\omega_0$. 
Looking stroboscopically at the dynamics (at integer multiples of the period) we see therefore a situation exactly equal to the Rabi oscillations: the beatings in Fig.~\ref{time_unitary:fig} occur at a frequency given by the quasi-energy difference. We conclude therefore that the condition~\eqref{adiabone:eqn} is not sufficient for getting adiabatic evolution in our system 
and it must be supplemented with $\omega_0$ being far from any quasi-degeneracy. 
Because all periodically driven systems show quasi-resonances where the Floquet states are exchanged we can conclude that this is a quite general statement, which can be applied also to more general periodically driven systems. \textcolor{black}{Our findings are in agreement with the results of Refs.~\cite{Gong1} obtained in the context of Thouless pumping: here an adiabatic perturbation theory is developed which gives rise to diverging denominators when the adiabatic quasi-energies are resonant.}

\section{Mapping to a two-band Wannier-Stark ladder}

It is instructive to analyse the mechanism behind the Floquet resonances in the spin-1/2 model by mapping this problem into that of a Wannier-Stark ladder.
As pioneered by Shirley~\cite{Shirley_PR65}, the Floquet modes and quasi-energies can be obtained by diagonalizing the Shirley-Floquet 
Hamiltonian, obtained by representing the operator $\hat{{\mathcal K}} = \hat{H}(t)-i\partial_t$ in the extended Hilbert space 
$\mathcal{H}\otimes\mathcal{R}$~\cite{Shirley_PR65,Sambe_PRA73}, where $\mathcal{H}$ is the Hilbert space of the Hamiltonian and $\mathcal{R}$ 
is generated by the basis $\left\{\nep^{-in\omega_0t}\right\}_{n\in{\bf Z}}$. 
The matrix for $\hat{{\mathcal K}}$ has a block form given by
%
\begin{equation} \label{hamiltoni:eqn}
 \big( \hat{{\mathcal K}} \big) = 
 \left(\begin{array}{cccccc}
        \ddots     &    \vdots        &       \vdots          &\\
        \hat{\mathrm H}_0-(n-1)\omega_0&\hat{\mathrm H}_1&\hat{\mathrm H}_2&\dots\\
        \hat{\mathrm H}_{-1}&\hat{\mathrm H}_0-n\omega_0&\hat{\mathrm H}_1&\dots\\
        \hat{\mathrm H}_{-2}&     \hat{\mathrm H}_{-1}& \hat{\mathrm H}_0-(n+1)\omega_0&\dots\\
         \vdots    &     \vdots       & \vdots     &  \ddots
        \end{array}\right)\,,
\end{equation}
%
where $\hat{\mathrm H}_n\equiv\frac{1}{\tau}\int_0^\tau\hat{H}(t)\,\nep^{in\omega_0t}\,\ud t$ is the $n$-th Fourier coefficient of the $\tau$-periodic operator $\hat{H}(t)$.
Notice that, when $\omega_0=0$ this can be seen as a model with one extra dimension --- the Fourier label $n$ --- and there is translation invariance, like in 
a tight-binding model: in this case $\hat{{\mathcal K}}$ has a {bounded} spectrum. 
Taking $\omega_0\neq 0$ is formally analogous to applying a uniform electric field to such a tight-binding model: the spectrum changes qualitatively 
its nature and becomes a Wannier-Stark-like ladder of doublets~\cite{mendez_PT81,Wannier:book} which we term ``Floquet-Stark ladder''. 
A genuine perturbative approach in $\omega_0$ is therefore impossible: the parameter $\omega_0$ makes the diagonal elements of the matrix unbounded 
(as $n\to \infty$) and qualitatively changes the nature of the spectrum.

The Shirley approach leads to a static eigenvalue problem determining Floquet quasi-energies and modes. 
We now introduce a scheme to study the {\em dynamics} of the problem. 
To simplify our formulas, we consider the totally equivalent Hamiltonian
\begin{equation} \label{h1:eqn}
  \hat{H}(t) = \Big( \epsilon + A \sin(\omega_0 t) \Big) \hat{\sigma}^z + \Delta \hat{\sigma}^x\,.
\end{equation}
%
%
obtained by modifying the phase of the periodic driving, and performing a rotation in spin space.
Our dynamical treatment relies on making the following {\em Ansatz} for the state of the system at time $t$ 
\footnote{This expansion is unique, as one can check by considering that $\ket{\Psi(t)}$ solves the Schr\"odinger equation with the time-periodic Hamiltonian Eq.~\eqref{h1:eqn}. 
Therefore, $\ket{\Psi(t)}$ can be expanded in a unique way in the Floquet basis; 
it is therefore written as a superposition with time-dependent coefficients of the $\tau$-periodic Floquet modes. 
Eq.~\eqref{pseudo_Fourier:eqn} is obtained by expanding each Floquet mode in Fourier series and collecting the time-dependent coefficient of each 
$\nep^{-in\omega_0 t}$ factor in the definition of $\ket{\psi_n(t)}$.}
\begin{equation} \label{pseudo_Fourier:eqn}
  \ket{\Psi(t)}=\frac{1}{\sqrt{\tau}} \sum_{n=-\infty}^\infty \ket{\psi_n(t)}\nep^{-in\omega_0 t}\,.
\end{equation}
This expression resembles a Fourier expansion, \textcolor{black}{but it is {\em not} a Fourier expansion because} the components $\ket{\psi_n(t)}$ have a generic dependence on time. 
The Shirley-Floquet eigenvalue problem is re-obtained when $|\psi_n(t)\rangle = \nep^{-i\mu_{\alpha} t} |\phi_{n,\alpha}\rangle$, where 
$|\phi_{n,\alpha}\rangle$ is the $n$-th component of the Fourier expansion of the Floquet mode $|\Phi_{\alpha}(t)\rangle$~\cite{Shirley_PR65,Sambe_PRA73,Grifoni_PR98}.
Substituting the expression for $\ket{\Psi(t)}$ in the Schr\"odinger equation with Hamiltonian Eq.~\eqref{h1:eqn}, 
and observing that  the block matrix in Eq.~\eqref{hamiltoni:eqn} is now tridiagonal, we obtain a nearest-neighbor tight binding form 
on a one-dimensional lattice (Floquet lattice) whose sites are labeled by the Fourier index $n$
\begin{equation} \label{tight_binding:eqn}
  i\partial_t\ket{\psi_n(t) }=\left(\epsilon\hat{\sigma}^z-n\omega_0+ \Delta\hat{\sigma}^x\right)\ket{\psi_n(t) } +
  \frac{A}{2i}\hat{\sigma}^z\left(\ket{\psi_{n+1}(t) }-\ket{\psi_{n-1}(t) }\right)\;.
\end{equation}
%
Although technically different, a mapping of a spin driven with two frequencies over a two-dimensional lattice with an effective electric field has also been performed in Ref.~\cite{Halperin_arxiv}.
As anticipated above, the problem in Eq.~\eqref{tight_binding:eqn} is a Wannier-Stark ladder for a two-band system, with $\omega_0$ playing the role of an electric field. 
%
%
%
We find convenient to apply a unitary transformation which rotates towards the basis of the $t=0$-eigenstates of the Hamiltonian Eq.~\eqref{h1:eqn}: 
\begin{equation} \label{rotat:eqn}
\ket{\widetilde{\psi}_n(t)} =\nep^{\frac{i}{2}\hat{\sigma}^y \atan(\frac{\Delta}{\epsilon})} \ket{{\psi}_n(t)} = 
\left( \begin{array}{l} u_n(t) \\ v_n(t) \end{array} \right) \;, 
\end{equation}
where $u_n$ and $v_n$ denote, respectively, the upper and lower band components of the resulting spinor. 
With these definitions, we can write the Schr\"odinger dynamics of our problem as:
\begin{equation} \label{FWS:eqn}
 \left\{\begin{array}{l}
  i\partial_t u_n(t) = \left(E-n\omega_0\right) u_n(t)  + \frac{A\epsilon}{2iE}\left( u_{n+1}(t)  - u_{n-1}(t) \right)
  -\frac{A\Delta}{2iE}\left( v_{n+1}(t)  - v_{n-1}(t) \right)\\
\\
  i\partial_t  v_n(t) = \left(-E-n\omega_0\right) v_n(t)  - \frac{A\epsilon}{2iE} \left( v_{n+1}(t)  - v_{n-1}(t) \right)
  -\frac{A\Delta}{2iE} \left( u_{n+1}(t) -  u_{n-1}(t) \right)
  \end{array}\right.
\end{equation}
%
%
where we have defined $E\equiv\sqrt{\epsilon^2+\Delta^2}$.
%
For $\omega_0=0$, the Floquet-lattice is homogeneous --- there is no electric-field-like tilting --- and we might associate to the problem
a translation-invariant two-band Hamiltonian $\hat{H}_0$. 
The corresponding energy bands can be easily calculated to have a dispersion
\begin{equation} \label{bandulae:eqn}
  \pm \epsilon_q=\pm \sqrt{\Delta^2+\left(\epsilon+A\sin(q)\right)^2} \hspace{5mm} {\rm with}\;q\in[0,2\pi]\,.
\end{equation}
For $\omega_0\neq 0$, the spectrum changes structure and becomes a ladder of doublets without bounds along the $n\to \pm \infty$ Floquet-lattice directions.
By preparing the system in a Bloch wave $\psi_q$ of the Hamiltonian $\hat{H}_0$, a standard semi-classical 
argument~\cite{Grosso:book}[Chap.1, Sec.6],~\cite{landau9}[Chap.6, Sec.56-57] shows that
the quasi-momentum $q(t)$ evolves linearly in time as $q(t)=q_0+\omega_0 t$. 
Moreover, if the particle is prepared in a Bloch wave-packet $|\Psi_{q(t)}\rangle$ around a given $q(t)$, 
the expectation value of the position on the lattice, $n(t)$, is given by  
\begin{equation}
\partial_t n(t) = \partial_q \left. \langle \Psi_{q} | \hat{H}_0 |\Psi_{q} \rangle \right|_{q=q(t)} \;. 
\end{equation} 
By explicitly calculating $\langle \Psi_{q} | \hat{H}_0 |\Psi_{q} \rangle$, using the {\em Ansatz} 
\begin{equation}
|\Psi_q\rangle = \sqrt{p_+(t)} |\Psi_{q,+}\rangle +\nep^{i\varphi(t)}\sqrt{p_-(t)} |\Psi_{q,-}\rangle 
\end{equation} 
in terms of Bloch wave-packets $|\Psi_{q,\pm}\rangle$ of energy $\pm \epsilon_q$, with $\varphi(t)$ an irrelevant phase factor,  
we finally arrive at the following semi-classical equations for the position $n(t)$ and the quasi-momentum $q(t)$: 
%
\begin{eqnarray} \label{permotus:eqn}
  \partial_t n(t) &=& (p_+(t)-p_-(t))\,\partial_q\epsilon_q\nonumber\\
  \partial_t q(t) &=&\omega_0\,.
\end{eqnarray}
Notice the appearance in our two-band problem of the quantities $p_-(t)=\sum_n |v_n(t)|^2$ and $p_+(t)=1-p_-(t)$, which are respectively 
the occupations of the lower and upper bands: these quantities are not determined by the semi-classical equations, but by the
microscopic equations for $v_n(t)$ and $u_n(t)$. We will discuss below their role and behaviour in the different regimes of interest.
The equation for $q(t)$, assuming $q_0=0$, is trivially solved by $q(t)=\omega_0 t$. 
The energy absorption, following Ref.~\cite{Halperin_arxiv}, is given by
\begin{equation} \label{derene:eqn}
  \partial_t E(t) = \omega_0 \,\partial_t n(t) =\omega_0 \, (p_+(t)-p_-(t)) \left.\partial_q\epsilon_q\right|_{q=\omega_0 t}\,.
\end{equation}
In the following, we prepare the system in the initial ground state ($p_-(0)=1$ and $p_+(0)=0$) and we study its energy evolution. 
According to the value of the frequency $\omega_0$, taken much smaller than $\min_{t\in[0,\tau]}E_{\rm gap}\left(t\right)$, we find two different regimes: 
the adiabatic one and the resonant one, corresponding to the two regimes discussed in Fig.~\ref{time_unitary:fig}, which we now illustrate in more detail. 
%
\subsection{Adiabatic regime}
We start assuming that the ``tilting'' $\omega_0$ is very small and far from any resonance: there is no level of the lower manifold which is degenerate with some level of the upper manifold. 
In these conditions, there is no excitation from the lower to the upper manifold, so $p_-(t)=1$ and $p_+(t)=0$. 
The validity of the adiabatic approximation can be directly seen in the extended Hilbert space: with small frequency there is a small tilting of the Floquet-Stark ladder, 
no levels are put in resonance and the Floquet lattice always stays in the lower band. 
Correspondingly, the physical system always stays in the ground state and the energy is always the ground state energy. 
To see this fact, we can evaluate the $q$-derivative in Eq.~\eqref{bandulae:eqn} and, using Eq.~\eqref{derene:eqn} and $p_-(t)-p_+(t)=1$, we obtain
\begin{equation} \label{eqn:E_t}
  \partial_t E =  \frac{2A\omega_0\left(\epsilon+A\sin\left(\omega_0 t\right)\right)\cos\left(\omega_0 t\right)}{\sqrt{\Delta^2+(\epsilon+A\sin(\omega_0 t))^2}}\,.
\end{equation}
We see that the energy periodically oscillates in time around a vanishing average: integrating we find that the energy is instantaneously equal to the adiabatic ground state value, 
$E(t)=-\sqrt{\Delta^2+\left(\epsilon+A\sin(\omega_0 t)\right)^2}$. 
The excitation energy is therefore vanishing: that is what we see in the low-frequency non-resonant case in Fig.~\ref{time_unitary:fig}. 
It is important to notice that the oscillations of the energy are the physical manifestation of the {\em Bloch oscillations}~\cite{Grosso:book} of $n(t)$ in the Floquet Wannier-Stark ladder Eq.~\eqref{FWS:eqn}.
\subsection{Resonant regime}
\begin{figure*}
  \begin{center}
    \begin{tabular}{c}
%
      \hspace{0cm}\resizebox{120mm}{!}{\includegraphics{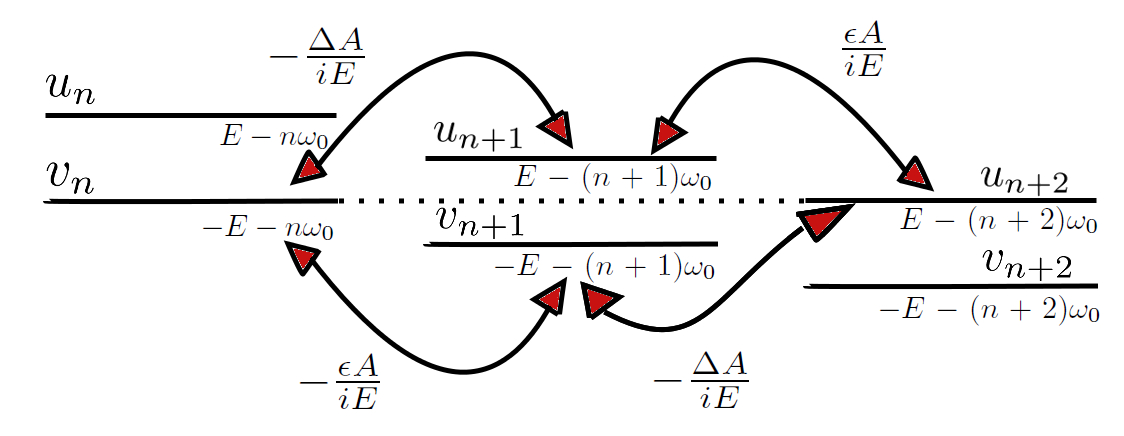}}\\
    \end{tabular}
  \end{center}
\caption{Second order resonance in the Floquet-Stark ladder (see main text).}
\label{fig1:fig}
\end{figure*}

Consider now the limit of small amplitude, $A\ll \{ \omega_0,\,\sqrt{\epsilon^2+A^2}\}$, and imagine that $\omega_0$ is such that there is an 
$m$-th order resonance: $E-m\omega_0=-E$. 
When the amplitude is small, this $m$-th order resonance corresponds to a Floquet quasi-degeneracy. 
Here adiabaticity is lost: we are going to see how this phenomenon can be interpreted in the Floquet lattice representation in the small amplitude limit. 
When the resonance condition is fulfilled, one level of the lower manifold is in resonance with some level of the upper one (Figure~\ref{fig1:fig} illustrates this situations for a resonance with $m=2$): the resulting Rabi oscillations 
between these two levels give rise to the wide energy oscillations observed in the lower panel of Fig.~\ref{time_unitary:fig}. 
We see that we can approximately restrict our dynamics to the decoupled subspaces formed by $v_n(t)$ and $u_{n+2}(t)$. 
Assuming $A\ll \omega_0$, we can apply the second order perturbation theory (as schematized in Figure~\ref{fig1:fig}): two amplitudes obey the equations
\begin{equation}
  \left\{\begin{array}{rl}
           i\partial_t v_{n}(t)     & \simeq  (-E +\delta E- n\omega_0) v_{n}(t) +\dfrac{2\epsilon\Delta A^2}{E^2\omega_0}u_{n+2}(t)\\
\\
           i\partial_t u_{n+2}(t) & \simeq  (-E -\delta E- n\omega_0) u_{n+2}(t) + \dfrac{2\epsilon\Delta A^2}{E^2\omega_0} v_{n}(t)
         \end{array}\right.\,,
\end{equation}
%
where, where $E=\sqrt{\epsilon^2+\Delta^2}$, $\delta E = \frac{(\epsilon^2-\Delta^2)A^2}{E^2\omega_0}$ and we have kept only the lowest perturbative order in $A/\omega_0$. 
This formula can be generalized to the case of resonance of order $m$: using again perturbation theory we get
%
%
\begin{equation}
   \left\{\begin{array}{rl}
           i\partial_tv_{n}(t)    & \simeq  (-E +\delta E- n\omega_0) v_{n}(t)
+J u_{n+m}(t)\\
\\
           i\partial_t u_{n+m}(t) & \simeq  (-E -\delta E- n\omega_0) u_{n+m}(t) 
+J v_{n}(t)
         \end{array}\right.\,,
\end{equation}
%
%
%
where $J,\,\delta E=\mathcal{O}(A^m/\omega_0^{m-1})$. 
%
%
Solving these equations, we find that the populations $p_-(t)$ and $p_+(t)$ undergo Rabi oscillations with frequency 
with frequency $\omega_L=\sqrt{(\delta E)^2+J^2}=\mathcal{O}(A^m/\omega_0^{m-1})$ (it is easy to see that $\omega_L=\frac{(\epsilon^2+\delta^2) A^2}{E^2\omega_0}$ when $m=2$).
So, in the evolution of the energy, see Eq.~\eqref{eqn:E_t}, we expect to see not only the fast oscillations of frequency $\omega_0$, but also slower oscillations of large 
amplitude and frequency $\omega_L$. 
Preparing the system in the ground state is equivalent to preparing the Floquet lattice in the lower manifold; the population imbalance $p_+(t)-p_-(t)$,  probed stroboscopically at times $t=n\tau$, periodically oscillates between -1 and $\frac{8\epsilon^2\Delta^2}{(\epsilon^2+\Delta^2)^2}$ with a period $2\pi/\omega_L$.
In the physical system this fact manifests in the energy moving from the ground state value to a maximum and then back, in a period $2\pi/\omega_L$. 
This is exactly what we see in the excitation energy in the lower panel of Fig.~\ref{time_unitary:fig}, for a value of $\omega_0$ where there is a quasi-degeneracy in the Floquet spectrum. 
Quite nicely the frequency $\omega_L$ corresponds to the perturbative splitting of the Floquet quasi-energies: as remarked above, the beatings occur with a frequency given by the 
Floquet gap at the quasi-degeneracy. 
Notice that these quasi-degenerate doublets occur when the frequency obeys the resonance condition $\omega_0=2E/m$, which provides a simple illustration 
for the fact that $\omega_0=0$ is a singular accumulation point for quasi-degeneracies.

\section{The effect of dissipation}
\begin{figure}
\begin{center}
   \includegraphics[width=8.3cm]{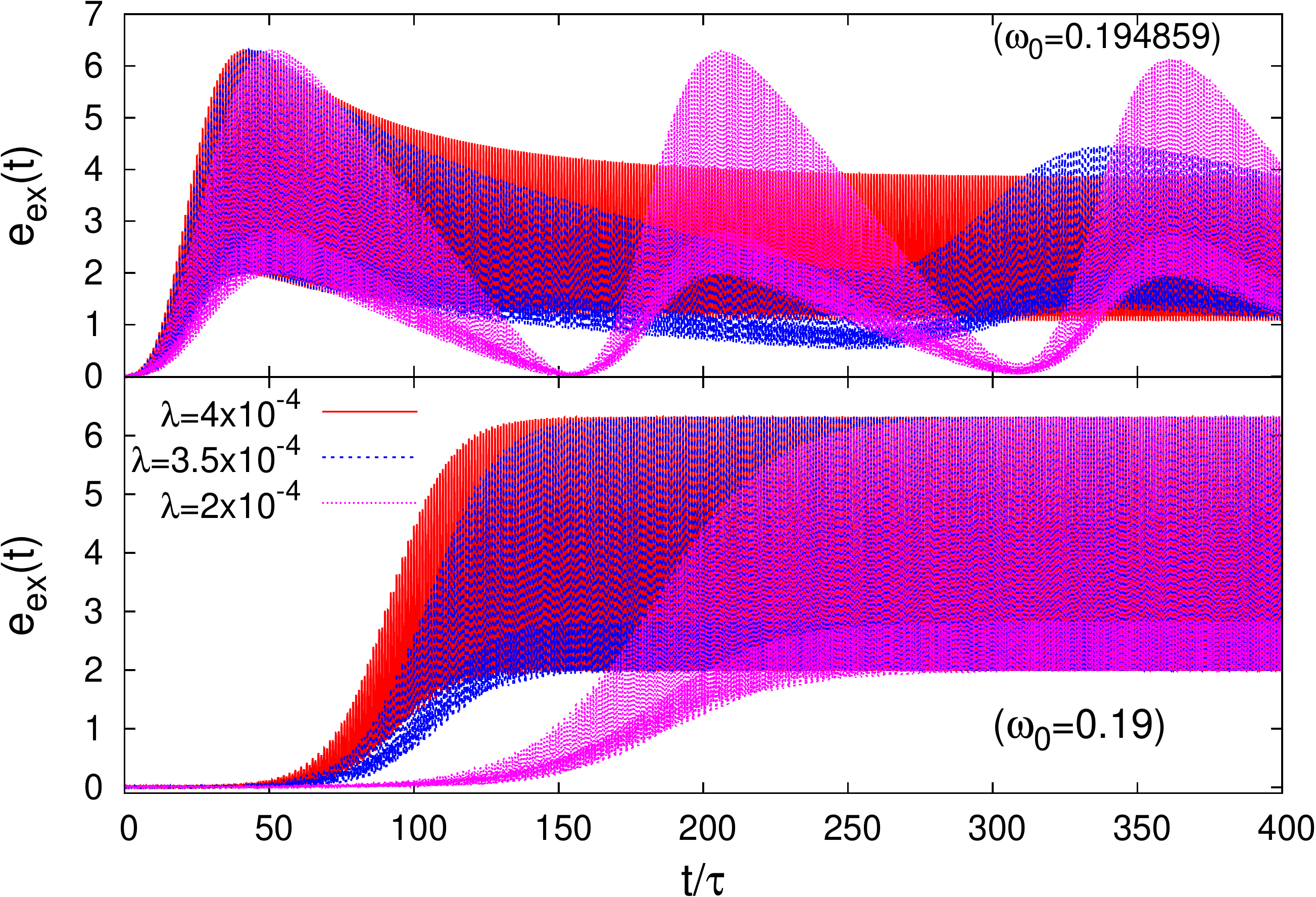}
\end{center}
\caption{(Upper panel) Landau-Lifshits-Gilbert evolution (Eq.~\eqref{llg:eq}) of the excitation energy for the classical precession equation 
for a closed-system quasi-degeneracy frequency ($\omega_0=0.194859$ -- upper panel) and a closed-system adiabatic one ($\omega_0=0.19$ -- lower panel). 
We take different values of $\lambda$: notice that when $\lambda$ is small enough, the large beatings at a Floquet quasi-degeneracy of the system without dissipation 
can be still observed (upper panel). (Numerical parameters: $\Delta=\epsilon=1,\,A=2$.)}
\label{LLG:fig}
\end{figure}

The crucial question that we now plan to tackle is to what extent the physics of the resonances seen before survives the effect of dissipation.
This is a crucial question, since experimentally, while long coherence times are indeed possible ---
for instance in superconducting qubits (see Refs.~\cite{Clarke_nat08,Devoret_sci13} for a review) ---, a perfect decoupling from the environment is impossible. 
In a classical setting of a bulk ferromagnet in a time-dependent magnetic field, one would introduce dissipation phenomenologically 
by supplementing the classical Eq.~\ref{class} with the Landau-Lifshits-Gilbert term \cite{Helmut:book,LLG:review}:
replacing $\mean{\hat{\boldsymbol{\sigma}}}_t$ with ${\bf M}(t)$ we would write
\begin{equation} \label{llg:eq}
  \frac{d}{dt}{\bf M}(t)=-2{\bf M}(t)\times \Big( {\bf B}(t) + \lambda {\bf M}(t) \times {\bf B}(t) \Big)\;,
\end{equation}
$\lambda$ being the Gilbert dissipation parameter. Notice that the modulus of ${\textbf M}(t)$ is conserved by this dissipative dynamics, while the energy is not.  
Fig.~\ref{LLG:fig} shows the results obtained at resonance ($\omega_0=0.194859$, top) compared to the off-resonance ones ($\omega_0=0.19$, bottom).
Away from resonance, the system gets excited above its instantaneous lowest-energy state in a time-scale which becomes longer as the dissipation parameter $\lambda$
decreases; the excitation energy shows $\tau$-periodic oscillations with an amplitude and an average that grows with time, until it settles to a stationary periodic regime. 
When the frequency $\omega_0$ is at resonance, on the contrary, the system shows beating-like oscillations for small values of $\lambda$, which become 
increasingly damped for growing $\lambda$. 
Notice that the amplitude of the asymptotic $\tau$-periodic oscillations is here smaller than that of the off-resonant case.
%
%

We now contrast such a classical picture with a fully quantum mechanical and microscopic one. 
We take into account the interaction of a single spin with an environment. 
Following Refs.~\cite{Vernon,Leggett_Caldeira_AP83} we model the environment as an Ohmic bath of harmonic oscillators at temperature $T=1/(k_B\beta)$. 
The total Hamiltonian is:
\begin{equation}
  \hat{H}_{\rm tot}(t) = \hat{H}(t) + \hat{H}_B + \hat{H}_{SB} \;,
\end{equation}
where we take $\hat{H}_B=\sum_n\left(\frac{\hat{p}_n^2}{2m}+\frac{1}{2}m\omega_n^2\hat{x}_n^2\right)$ and assume 
$H_{SB}=\sum_n k_n \hat{x}_n\otimes\hat{\sigma}^z$. 
Using the Bloch-Redfield approximation \cite{Breuer:book,Cohen:book,Blum:book} we can write a closed equation of motion for 
the reduced density matrix of the system. 
In the interaction representation with respect to $\hat{H}(t)$ we have: 
%
\begin{eqnarray}  \label{master}
  \dot{\rho}_I(t) = - 
  \left( \int_{0}^{+\infty}\!\! \ud s \, C(s) \, \big[ \sigma_I^z(t), \sigma^z_I(t-s) \rho_I(t) \big] + {\rm H.c.} \right) \;,
\end{eqnarray}
where $C(s)$ denotes the bath correlation function 
%
\begin{equation}
C(s) = \sum_n k_n^2 \langle \hat{x}_n(s) \hat{x}_n(0) \rangle = 
\int_0^{+\infty} \!\! \ud \omega \; J(\omega) \Big( \nep^{i\omega s} \mathcal{N}(\omega) +  \nep^{-i\omega s} (\mathcal{N}(\omega)+1) \Big) 
\end{equation}
which is in turn expressed in terms of the spectral density $J(\omega)\equiv\sum_n\frac{k_n^2}{2m_n\omega_n}\delta(\omega-\omega_n)$, 
and of the Bose occupation factor $\mathcal{N}\left(\omega\right)=\frac{1}{\nep^{\beta\omega}-1}$. 
In the following, we will assume, for the spectral density, the Ohmic form~\cite{Leggett_Caldeira_AP83} 
$J(\omega)=\frac{2\gamma}{\pi}\omega\frac{\Omega^2}{\omega^2+\Omega^2}$, where $\Omega$ is a regularizing Lorentzian cutoff. 
To proceed, we expand Eq.~\eqref{master} in the basis of the Floquet modes at time 0, $\left\{\ket{\Phi^+(0)},\,\ket{\Phi^-(0)}\right\}$ and apply the 
Rotating Wave Approximation (RWA) in this basis~\cite{Russomanno_PRB11,Koheler_PRE97,Hausinger_PRA10}. 
Going back to the Schr\"odinger representation, we find that the density matrix in the basis of the instantaneous Floquet modes 
$\left\{\ket{\Phi^+(t)},\,\ket{\Phi^-(t)}\right\}$ relaxes after a transient to a time-independent diagonal matrix whose populations, 
$\rho_{++}^{\rm eq}$ and $\rho_{--}^{\rm eq}=1-\rho_{++}^{\rm eq}$, are given by~\cite{Russomanno_PRB11}:
%
\begin{eqnarray} \label{steady}
  \rho_{++}^{\rm eq} &=& \frac{\displaystyle \sum_{n=-\infty}^{+\infty} J\left(n\omega_0+2\mu\right) \mathcal{N}\left(n\omega_0+2\mu\right) \left| \sigma^{z}_{-n} \right|^2}
  {\displaystyle \sum_{n=-\infty}^{+\infty} J\left(n\omega_0+2\mu\right) \Big(2\mathcal{N}\left(n\omega_0+2\mu\right) +1\Big) \left| \sigma^{z}_{-n}\right|^2} \;.
\end{eqnarray}
Notice that we use here that $J(-|\omega |)=-J(|\omega |)$ and $\mathcal{N}(-\omega)=-(\mathcal{N}(\omega)+1)$. 
We also defined $\sigma^{z}_n$ as the $n$-th Fourier coefficient 
\begin{equation}
\sigma^z_n=\frac{1}{\tau} \int_0^\tau \ud t \; \nep^{in\omega_0 t}  \bra{\Phi^+(t)}\hat{\sigma}^z\ket{\Phi^-(t)} \;.
\end{equation}
We observe that the density matrix in the long-time limit is periodic, because it is constant in the periodic basis $\left\{\ket{\Phi^+(t)},\,\ket{\Phi^-(t)}\right\}$: 
it reaches a periodic steady state in which all the observables become periodic, even the excitation energy. 
In this regime, the system is stationary: in each cycle it absorbs from the forcing field as much energy as it gives to the thermal bath. 

\begin{figure}
\begin{center}
   \includegraphics[width=8.3cm]{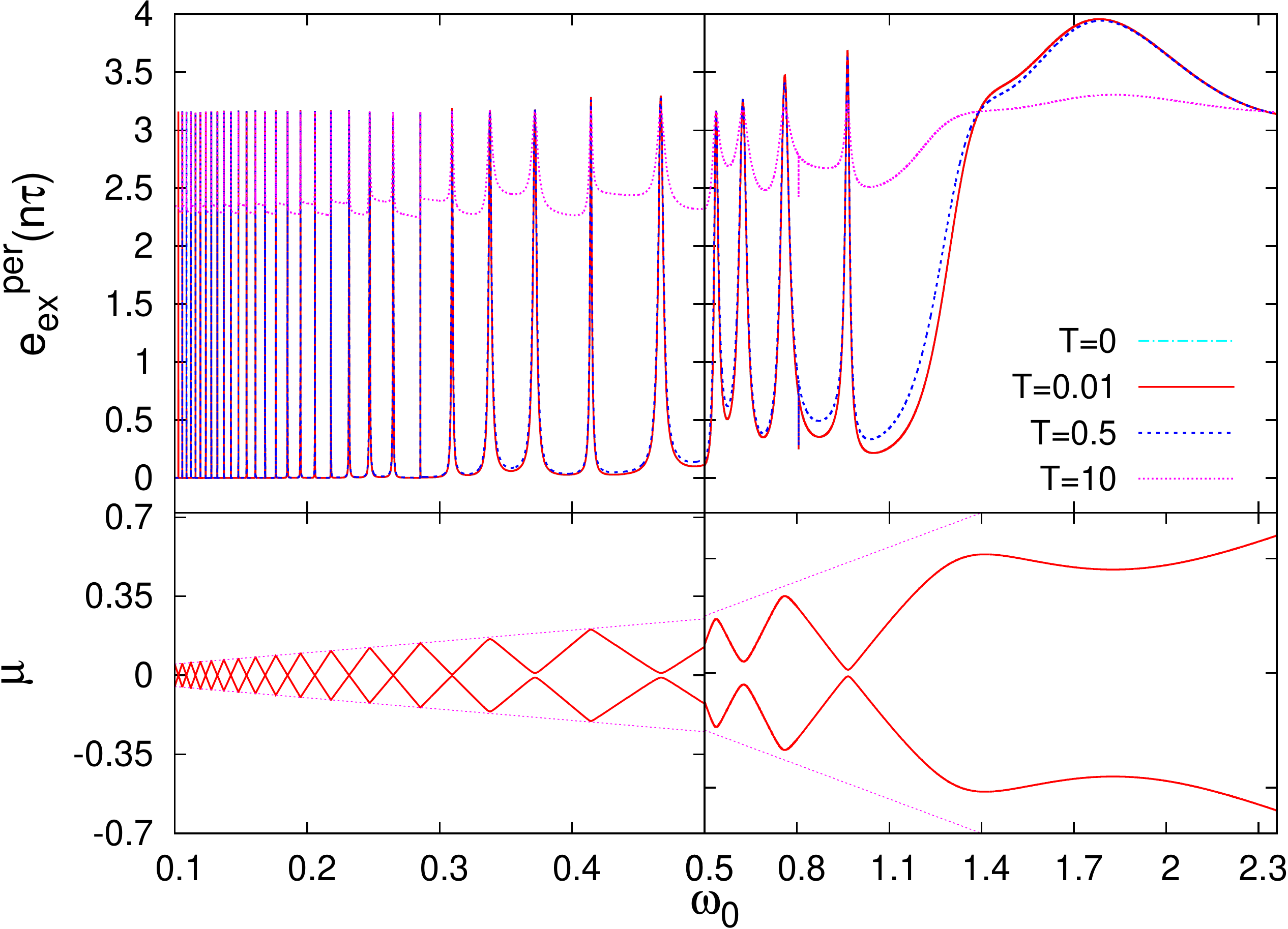}
\end{center}
\caption{(Upper panel) Excitation energy in the periodic steady state at the end of a period vs. $\omega_0$ in the dissipative case; 
we used Eq.\eqref{steady} for different values of the temperature $T$. 
Notice the peaks occurring at the frequencies $\omega_0$ where the Floquet quasi-degeneracies occur (lower panel). 
Notice that the cases for $T=0$ and $T=0.01$ are numerically indistinguishable. 
(Numerical parameters: $\Delta=\epsilon=1,\,A=2,\,\Omega=500$; the results are independent of the strength of the coupling to the environment $\gamma$, as long as small.)
}
\label{floq_noise_vs_om:fig}
\end{figure}

In Fig.~\ref{floq_noise_vs_om:fig} we plot the excitation energy in the periodic steady state at the end of a period, $e_{\rm ex}^{\rm per}\left(n\tau\right)$, versus $\omega_0$. 
We observe peaks in the stationary excitation energy exactly at the Floquet quasi-degeneracies: 
when $\omega_0$ is at a quasi-degeneracy, the energy absorption of the system 
is more efficient, even when we account for the dissipation towards an environment. 
In the low-frequency regime the quasi-degeneracies become infinitely dense and thin, and it becomes increasingly hard to observe the peaks numerically.
The behaviour away from the peaks is equally interesting: at low temperatures the excitation energy vanishes, exactly as in the unitary adiabatic case. 
At high temperatures, on the opposite, the excitation energy is different from 0 also far from any quasi-degeneracy, because of the thermal excitations,
and an increasing function of $T$. 
%

%
\section{Conclusion} In conclusion, we found that the adiabatic limit $\omega_0\to 0 $ is singular in its being ``decorated'' by interesting quasi-degeneracies 
in the Floquet spectrum, which give rise to a deviation from adiabaticity. Its physical consequences --- most strikingly a large increase in dissipation --- remarkably survive in presence of dissipation towards an environment. 
We can interpret the phenomenon in terms of Rabi oscillations of degenerate levels in the Floquet-Stark ladder in the extended Hilbert space representation. 
Perspectives of future work will focus on the interpretation of the dynamical localization phenomenon~\cite{Casati:rotor,Boris:rotor,Delande_PRL08} and the Thouless topological pumping~\cite{Thouless_PRB83} in terms of this formalism.
\ack{We acknowledge fruitful discussions with L.~Privitera and with Sir M.V. Berry. A.~R. acknowledges financial support from the EU integrated project QUIC, from his parents and from ``Progetti interni - Scuola Normale Superiore''; G.~E.~S. acknowledges financial support from ERC MODPHYSFRICT. This work was not supported by any
military agency.}
%

\end{document}